\documentclass[sigplan,screen,10pt]{acmart}
 
\acmConference[WoSC20]{WoSC20: Workshop on Serverless Computing}{Dec 07--11, 2020}{Delft, Netherlands}
\acmBooktitle{WoSC20: Workshop on Serverless Computing,
  Dec 07--11, 2020, Delft, Netherlands}

\usepackage{algorithm}
\usepackage{algpseudocode}
\usepackage{wrapfig}
\usepackage{xspace}
\usepackage{verbatim}

\newcommand{\tightparagraph}[1]{\vspace{5pt}\noindent\textbf{#1}\ }
\newcommand{\eg}{{\it e.g., }}
\newcommand{\ie}{{\it i.e., }}

\newcommand{\freshen}{${\texttt{freshen}}$\xspace}

\title{Proactive Serverless Function Resource Management}
\author{Erika Hunhoff, Shazal Irshad, Vijay Thurimella, Ali Tariq, Eric Rozner}
\date{Oct 2020}

\begin{document}

\begin{abstract}

This paper introduces a new primitive to serverless language runtimes called \freshen. With \freshen, developers or providers specify functionality to perform before a given function executes. This proactive technique allows for overheads associated with serverless functions to be mitigated at execution time, which improves function responsiveness. We show various predictive opportunities exist to run \freshen within reasonable time windows. A high-level design and implementation are described, along with preliminary results to show the potential benefits of our scheme.
\end{abstract}

\settopmatter{printfolios=true} 
\maketitle

\section{Introduction}
\label{sec:intro}

Serverless computing is an emerging paradigm in which cloud providers seamlessly scale developer-provided functions as demands change. Although seemingly simple, serverless functions have been shown to support a wide variety of workloads, from chat bots, video processing, machine learning, HCI, to even general compute. As serverless ecosystems mature, functions will be integrated into a set of larger and larger microservices and will also be relied upon to directly interface with users. As such, the execution latency of serverless functions becomes an important consideration.

However, the simplicity of today's serverless deployments may increase execution times. Consider a simple function, $\lambda_1$, which downloads a machine learning model from a server, analyzes an input image, and performs additional processing before writing a result to a datastore. Without special care, many overheads exist. After starting, the function must create a connection to the server hosting the model and then download the model from the server. This behavior could happen anew for subsequent instantiations of $\lambda_1$, even if  running sequentially in the same warmed container. When writing the result, another connection must be established before the data is sent. Again, this overhead could reoccur for each successive invocation of $\lambda_1$. These per-invocation overheads (\ie establishing connections, refetching the model, incurring TCP slow start, etc.) quickly add up, which is problematic because many functions have short execution times.

To deal with such issues, developers can utilize {\em runtime reuse.} In runtime reuse variables can be {\em runtime-scoped} inside the language runtime executing within the container the serverless function runs in.\footnote{We use ``container'' to generally refer to VMs or containers} Runtime-scoped variables can be accessed across subsequent serverless function instantiations within a given runtime and container. Revisiting our example, network connections can be reused within a runtime when defined as a runtime-scoped variable to avoid per-instantiation connection overheads.

In this paper, we argue runtime reuse is insufficient to overcome many of the redundant overheads described earlier. Even with runtime reuse, fetched data could be out-of-date, connections may revert their congestion windows to small initial values or even time out, or application-level data could be stale from the last invocation. To combat these issues, we propose a new primitive called \freshen{}, which can be {\em proactively} invoked by the serverless infrastructure. A \freshen hook is implemented within the runtime, allowing developers or providers to establish or warm connections, proactively fetch data, or otherwise perform actions to reduce overheads when the serverless function runs. The \freshen hook is designed to be run before its corresponding function is instantiated, and we contend this is possible because there are many opportunities to predict a function's instantiation before it is invoked. 

This paper provides motivation and background in Section \ref{sec:background}, a preliminary design in Section \ref{sec:design}, and potential benefits of \freshen in Section \ref{sec:eval}. Related Work is detailed in Section~\ref{sec:related}. Finally, a discussion and conclusion section is presented in Section~\ref{sec:discuss}.

\section{Background and Motivation}
\label{sec:background}
This section provides background on runtime reuse and highlights scenarios where runtime reuse may be inefficient. Then, we motivate ways to predict function instantiations.

\tightparagraph{Serverless runtime reuse}
While all providers allow runtime reuse, here we explain how an open-source platform, OpenWhisk, enables reuse. 
OpenWhisk runs functions within Docker containers, listening as a daemon on port 8080. 
After the Docker container is initialized, the \texttt{init} hook starts the language runtime within the container and also loads the actual function code. When the \texttt{run} hook is invoked, the function will be scheduled to run. 
Thus, the persistent runtime invoked during {\tt init} can be thought of as a program that listens for the {\tt run} hook, executes the function, and returns the result.

Without runtime reuse, variables are scoped for use within a single invocation only, which we term {\em invocation-scoped}. In contrast, {\em runtime-scoped} variables variables can be reused across serverless function instantiations in a given runtime. Common use cases for {\em runtime-scoped} variables are persistent network connections (ensuring connection quotas are not exhausted) and frequently-accessed data fetched during the first function invocation and then stored in the runtime for the lifetime of the container.

\begin{figure}[ht]
\begin{minipage}[b]{0.48\columnwidth}
\centering
 \includegraphics[width=1\linewidth, height =0.5\columnwidth]{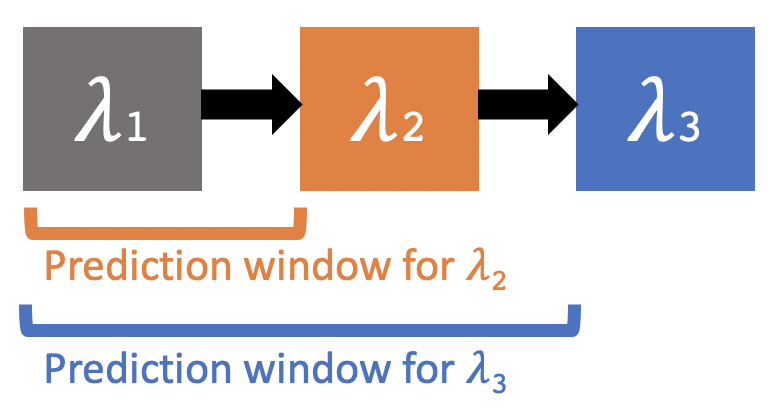}
 
    \captionsetup{font={small}}
    
   \captionof{figure}{ Opportunities for {\tt freshen} within a function chain}
   
    \label{fig:lambdachain}
    
\end{minipage}
\hspace{0.1cm}
\begin{minipage}[b]{0.48\columnwidth}
\centering
\includegraphics[width=1\linewidth, height =0.7\columnwidth]{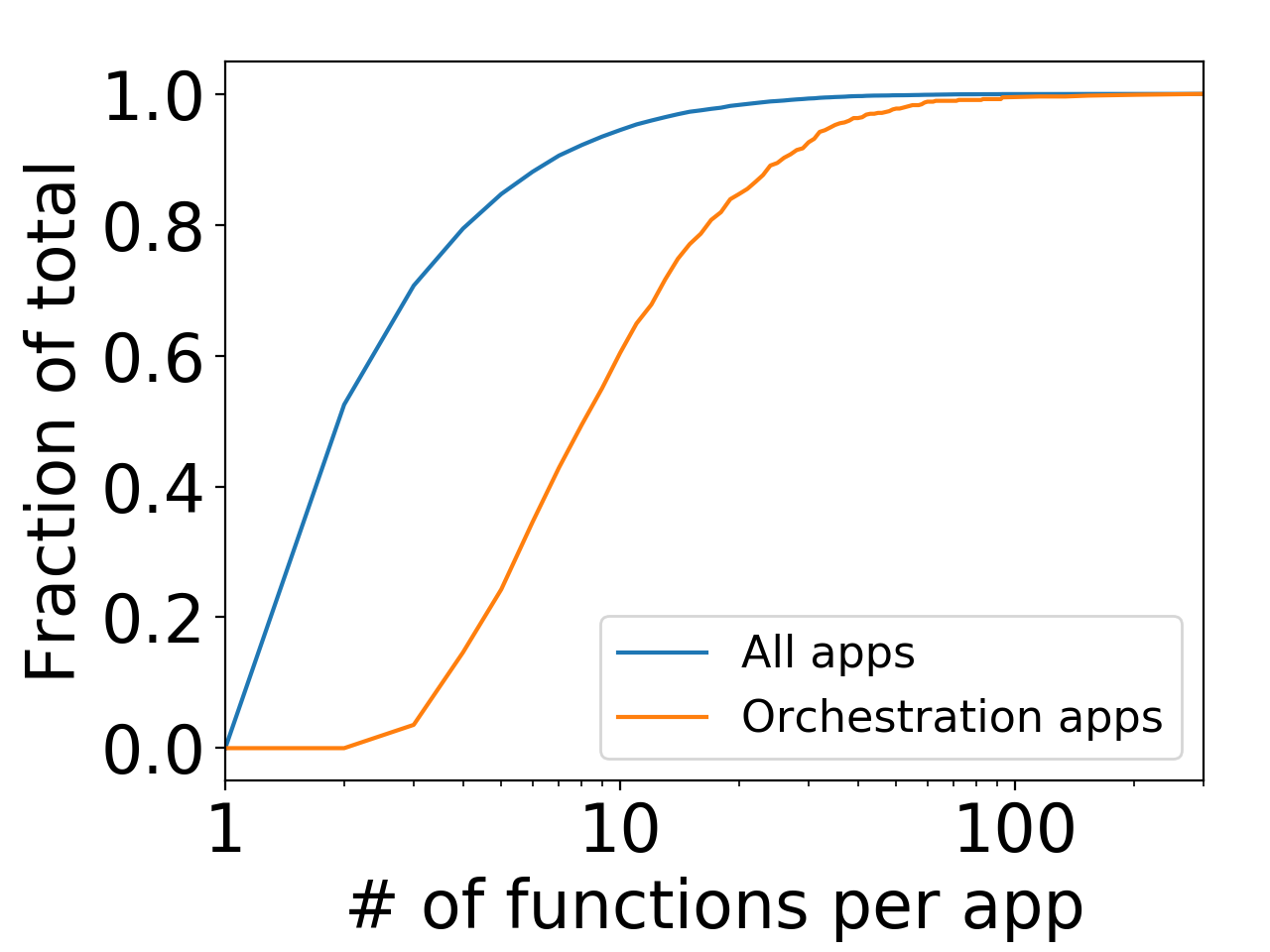}
\captionsetup{font={small}}
\caption{Orchestration apps have more functions in chains}
    \label{fig:azure-chain}
\end{minipage}

\end{figure}
\tightparagraph{Runtime reuse inefficiencies}
While runtime reuse can be used by application developers to increase application efficiency, numerous issues may arise. First, there may exist cases in which the runtime has not been initialized, such as when cold starts occur, meaning reuse is not possible. Some studies have shown inefficient container reuse across function invocations, which increases cold start frequency~\cite{sequoia}. Other studies indicate some serverless infrastructures
disallow container sharing between functions, which can also increase cold starts
when container resources are limited~\cite{curtains}. Second, there may be cases when the runtime is initialized, but the data held within the runtime is stale. For example, an object stored within the runtime may need to be retrieved from a datastore because a newer version is available. Network connections may have timed out or have reset their TCP state (\eg{} congestion window, round trip times, etc). The Linux congestion control algorithm reduces the congestion window (CWND) on inactive connections. Last, approaches to reduce connection (re)establishment overheads may not apply. The Linux  {\tt tcp\_no\_metrics\_save} capability allows metrics like RTT and ssthresh to be cached between TCP connections to the same destination, but does not apply to important parameters such as CWND. TCP Fast Open requires sender/receiver support and limits the amount of data sent in initial handshakes to small amounts. As a result, we believe several inefficiencies remain, even with runtime reuse, that can be addressed with {\tt freshen} called proactively before function instantiations.

\tightparagraph{Regaining efficiency via prediction} To alleviate the above concerns, we introduce a {\tt freshen} hook into the runtime, which can be called before a function is set to run. The {\tt freshen} hook allows providers or developers to execute arbitrary code intended to speed up function execution times. {\tt freshen} could warm pre-existing network connections, ensure locally-cached items are up-to-date, or even proactively retrieve a needed object. In order for {\tt freshen} to be effective, we must be able to predict when a function may run. There are several cases in which prediction may be possible. First, in serverless function chains, such as in Figure~\ref{fig:lambdachain}, explicit knowledge of a serverless function chain could predict impending function invocations within the chain. Function chains are often explicitly provided (as in Orchestration frameworks like AWS Step Functions) or can be derived via tracing or service mesh techniques~\cite{Mace:2018:UCP:3190508.3190526}.
To better understand prediction opportunities, we briefly study function chains in Orchestration frameworks. Figure~\ref{fig:azure-chain} shows a CDF of the number of functions within a single serverless application for Orchestration applications on Azure (data from traces in~\cite{serverless_in_the_wild}), compared to the number of functions within a single application over all applications. Orchestration frameworks are specifically designed to support function chains, and hence applications utilizing Orchestration frameworks typically consist of more functions: 8 functions in the median Orchestration case versus 2 functions in the median case of all. Considering a median function runtime of \textasciitilde{}700ms~\cite{serverless_in_the_wild},
opportunities for prediction could be as high as \textasciitilde{}5.6s in the extreme case of a linear chain dependency (as in Figure~\ref{fig:lambdachain}).

\begin{wraptable}{R}{0.50\columnwidth}
\small 
\centering
\begin{tabular}{| c | c |} 
 \hline
Trigger Service &  Delay (s) \\  \hline
Step Functions & 0.064 \\ \hline
Direct (Boto3) &  0.060  \\ \hline
SNS Pub/Sub &  0.253  \\ \hline
S3 bucket &  1.282  \\ \hline
\end{tabular}
\caption{Trigger overhead}
\label{tab:overhead-summary}
\end{wraptable}
In addition, functions within chains may be triggered by other services, such as a storage trigger, a pub/sub trigger, or a direct invocation. Table~\ref{tab:overhead-summary} shows the median delay, over 20k runs, between invoking a function via the listed service and the actual subsequent triggered function start time in AWS. Cold starts are carefully avoided, and the methodology in~\cite{sequoia} is used to obtain overheads by measuring timestamps just before the function trigger and at the start of the triggered-function. The table shows latencies range from 60ms to 1.28s, allowing time for the previous function within the chain, or the serverless provider, to call and execute \freshen on the next function within the chain.

\section{Design and Implementation}
\label{sec:design}

The following sections address when a \freshen hook could run (Section \ref{ssec:whenfresh}), what a \freshen hook could do (Section \ref{ssec:whatfresh}), and how \freshen could be implemented (Section \ref{ssec:freshimpl}).
Throughout, we will refer to $\lambda$, an example serverless function shown in pseudocode in Algorithm \ref{lambda}, to illustrate how \freshen could warm a connection and prefetch data. $\lambda$ first fetches some data (\texttt{DataGet}) over a network connection, performs some calculation based over the fetched data and the $\lambda$'s parameters, writes an output value to an external resource (\texttt{DataPut}), and finally returns whether the write was successful or not.

\begin{algorithm}
\caption{Sample Serverless Function $\lambda$}\label{lambda}
\begin{algorithmic}[1]
\State \text{Runtime Constants: } $CREDS, \, ID_1, \, ID_2$
\Procedure{$\lambda$}{$args$}
\State $data \coloneqq$ \textbf{DataGet(} $CREDS, \, ID_1$ \textbf{)}
\State $...$
\State $result \coloneqq ...$
\State $...$
\State $ret \coloneqq$ \textbf{DataPut(} $CREDS, \, ID_2, \, result$ \textbf{)}
\State \textbf{return} $ret$
\EndProcedure
\end{algorithmic}
\end{algorithm}

\subsection{When to \emph{freshen}}
 \label{ssec:whenfresh}
\begin{figure}
\begin{center}
\includegraphics[width=1\linewidth]{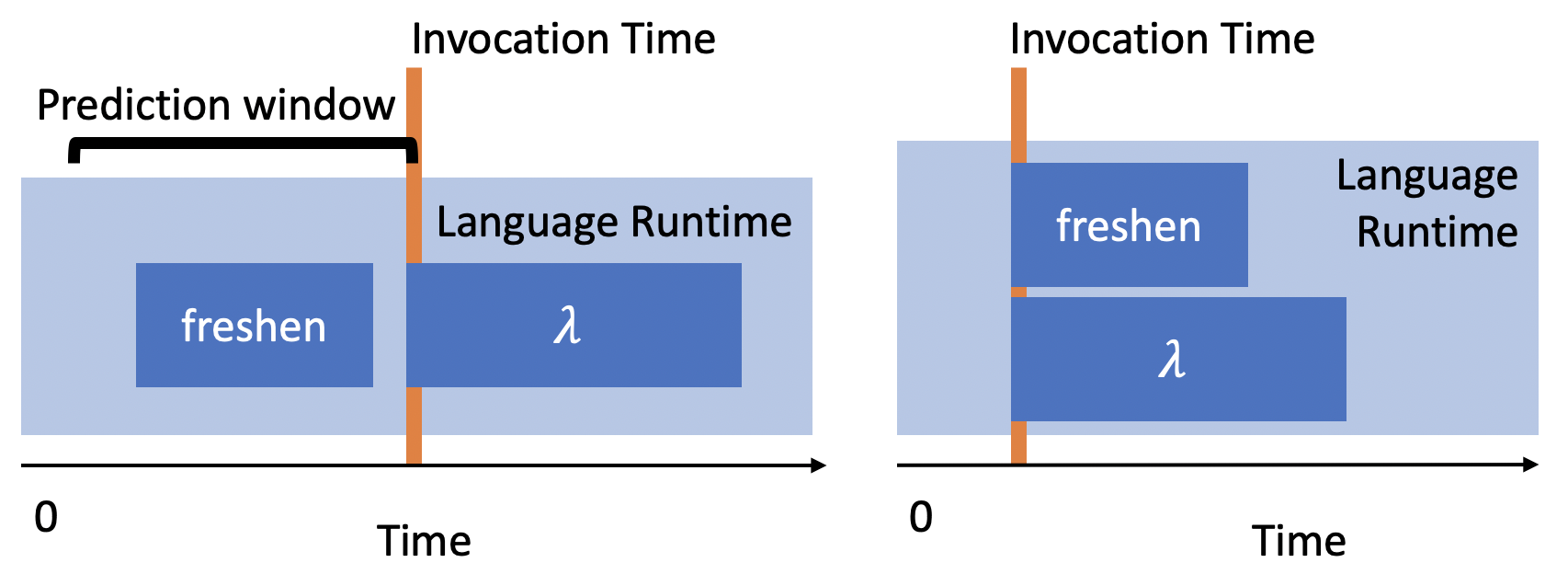}
\end{center}
\caption{\label{fig:freshentiming}Predicted (left) and unanticipated (right) timing of \freshen}
\end{figure}

The serverless framework would attempt to run \freshen before the serverless function (best case) or simultaneously (worst case). \freshen would be non-blocking and run within a separate thread in the language runtime so the logic and timing of function invocation via the \texttt{run} hook is unmodified. Figure~\ref{fig:freshentiming} shows the two examples of when \freshen could run in relation to the $\lambda$ it is freshening.

\subsection{Opportunities to \emph{freshen}}
\label{ssec:whatfresh}

\freshen could perform a variety of actions, including TCP connection establishment, 
TCP connection warming, 
state maintenance of other connection-oriented protocols, 
and proactive data fetching. 

\tightparagraph{Connection establishment and checks}
\label{sssec:tcpfreshen}
If a serverless function uses a resource with an underlying TCP connection, the function developer can either establish a runtime-scoped socket connection to take advantage of runtime reuse or create the connection as an ephemeral invocation-scoped variable. In both cases, \freshen could help reduce function latency. If the connection is runtime-scoped, \freshen would send a TCP keepalive to ascertain connection liveness; if the connection is not alive, \freshen could reestablish the connection. If the connection is invocation-scoped, \freshen could proactively establish the connection before the function attempts to create it. 

\freshen would only be able to perform connection establishment for connections with constant arguments (\eg constant IP and port). We posit this is often the case, as serverless functions often interact with known services such as storage.

\tightparagraph{Connection warming}
\label{sssec:warmfreshen}
\freshen could also take steps to warm TCP connections used by the serverless function, for instance setting the CWND. This could be facilitated via a new system call, {\tt warm\_cwnd}, which would determine an appropriate value of CWND based on current network conditions and anticipated workloads. The CWND can be estimated via techniques like packet pair probing to determine the current bandwidth~\cite{keshav1995packet} or analyzing the CWND of recent similar TCP connections to the same destination. Repetitive invocations can be used to anticipate workload characteristics, which could guide the warming function on whether warming is appropriate.
The {\tt warm\_cwnd} function can set initial congestion windows or alter congestion windows on longer-running, inactive connections. Since {\tt warm\_cwnd} is implemented as a system call, final determination of actual CWND values, as well as  permissions on whether such values can be altered, resides within the provider who is running the underlying host infrastructure.

\tightparagraph{Other connection-oriented protocols}
\label{sssec:tlsfreshen}
\freshen can establish and warm other connection-oriented protocols and protocols that run on top of TCP such as TLS, as long as the credentials are constant. However, for TLS establishment and other user-space protocols, the serverless provider would require some knowledge of the libraries in order to create provider-generated \freshen hooks for those resources. Developers who write their own \freshen hooks, as detailed in Section~\ref{ssec:freshimpl}, would have access to such knowledge.

\tightparagraph{Proactive data fetching}
\label{sssec:cachefreshen}
Consider the $\lambda$ in Algorithm \ref{lambda}: if the data fetched with \texttt{DataGet} is retrieved using constant credentials and resource identifiers, it is possible to prefetch the data before $\lambda$ is invoked. 

Prefetching leads to the concept of a \freshen-maintained cache of prefetched data. If the function is invoked frequently within the same runtime and accesses a read-only data resource, it may only be necessary to fetch the data once every $n$ seconds instead of every time the function is run, reducing network traffic. The time-to-live (TTL) of values within the \freshen cache could be set by a default value, by \freshen configuration values specified by the function developer, or by modifying the \texttt{DataGet} library to configure the TTL value on a per-resource level. In the more general case, associated timestamps or version numbers could be used to determine the freshness of items in the runtime \freshen cache, and data could be updated the next time \freshen or the serverless function is called.

\subsection{Implementation}
\label{ssec:freshimpl}

In the simplest implementation of \freshen, the function developer would write \freshen for each serverless function that requires optimization. This would provide the most opportunity for customized optimization. As an interesting alternative, for common resources and for popular serverless languages (\eg JavaScript, Python) \freshen code could be inferred by the serverless framework itself.

Code generation would be complex but here we rely on several observations about serverless functions and frameworks to reduce the scope of the problem:
\begin{itemize}
\item If \freshen were unable to be inferred, the serverless framework could continue unmodified with no major performance loss. Hence, failure to infer is not fatal.
\item Source code is available for static analysis for such tasks as identification of read-only data fetched using constant parameters.
\item Identical function code is run multiple times, so dynamic tracing of functions to identify commonly accessed resources is possible (similar to the tracing technique presented in \cite{containerless}).
\item The latency cost of the network operations \freshen seeks to optimize are much slower than CPU speeds so some overhead for learning how to infer \freshen is permissible.
\item Implementing inference only for libraries used to access other cloud services offered by the serverless provider has the potential to lower latency for a majority of functions without having to infer \freshen behavior for unknown resources.
\end{itemize}

One option for implementing \freshen for scripting languages is to use added runtime-scoped state and dynamically-inserted wrapper functions. The purpose of the runtime-scoped state is to track and coordinate \freshen resources between the \freshen call and the actual function invocation. The purpose of the dynamically-inserted wrappers is to intercept access to freshened resources. We will illustrate a simplified example of what an inferred \freshen could resemble for $\lambda$ in Algorithm \ref{lambda}.

The runtime-scoped state would minimally be a collection of ordered \freshen resources. A {\em freshen resource} is any object or resource that the \freshen code may interact with, such as a socket or a data object. In our example, the \freshen resources are kept in an ordered runtime-scoped list called $fr\_state$. In Algorithm~\ref{lambda}, the \texttt{DataGet} operation which \freshen can fetch or prefetch, will be assigned index 0 since it is the first resource accessed by $\lambda$. \texttt{DataPut}, which \freshen can warm, is assigned index 1. Each entry in \texttt{fr\_state} could contain a variety of metadata, such as a \emph{state} (\eg \texttt{running}, \texttt{finished}, etc.), a \emph{result} (\eg the prefetched data), a \emph{TTL} for the result, and a \emph{timestamp} recording the last time that entry was freshened. For simplicity, we only consider \emph{state} and \emph{result} in the following algorithms.

\begin{algorithm}
\caption{Freshen Function for $\lambda$}\label{freshen}
\begin{algorithmic}[1]
\State \textbf{Runtime State: } $fr\_state$
\Procedure{$Freshen$}{}
\State $fr\_state{[0]} \coloneqq running$
\State $fr\_state{[0]}.result \coloneqq$ \textbf{DataGet(} $CREDS$, $ID_1$ \textbf{)}
\State $fr\_state{[0]} \coloneqq finished$
\State $fr\_state{[1]} \coloneqq running$
\State \textbf{DataPut.warm(} $CREDS$ \textbf{)} \label{freshen:datawarm}
\State $fr\_state{[1]} \coloneqq finished$
\State \textbf{return}
\EndProcedure
\end{algorithmic}
\end{algorithm}

Algorithm \ref{freshen} illustrates an example \freshen function for $\lambda$. As mentioned, {\tt DataGet} is assigned to index 0 and {\tt DataPut} is assigned to index 1. The states \emph{running} and \emph{finished} surround the {\tt DataPut} and {\tt DataGet} calls of \freshen, and are used to coordinate the execution of \freshen with the execution of $\lambda$. Algorithm \ref{annotatedlambda} is the annotated version of Algorithm \ref{lambda}. The function wrappers appear at lines~\ref{annotatedlambda:FrFetch} and~\ref{annotatedlambda:FrWarm}. The function wrappers used are {\tt FrFetch} (for \emph{freshen fetch}) and {\tt FrWarm} (for \emph{freshen warm}).

\begin{algorithm}
\caption{Annotated Sample Serverless Function}\label{annotatedlambda}
\begin{algorithmic}[1]
\State \text{Runtime Constants: } $CREDS, \, ID_1, \, ID_2$
\Procedure{$\lambda$}{$args$}
\State $data \coloneqq$ \textbf{FrFetch(} $0,$ \textbf{DataGet(} $CREDS$, $ID_1$ \textbf{))} \label{annotatedlambda:FrFetch}
\State $...$
\State $result\coloneqq ...$
\State $...$
\State $ret \coloneqq$ \textbf{FrWarm(} $1,$ \textbf{DataPut(} $CREDS, \, ID_2, \, result$ \textbf{))} \label{annotatedlambda:FrWarm}
\State \textbf{return} $ret$
\EndProcedure
\end{algorithmic}
\end{algorithm}

Algorithms \ref{freshfetch} and \ref{freshwarm} are the implementations of those wrappers. The main function of each wrapper is to synchronize \freshen actions with $\lambda$'s use of that resource. If the resource has already been freshened, the wrapper returns either the prefetched data (line~\ref{freshfetch:return1} in Algorithm \ref{freshfetch}) or nothing where \freshen's only job is to warm the resource (line~\ref{freshwarm:return1} in Algorithm \ref{freshwarm}). In Algorithm \ref{freshwarm} it is assumed that there is already some knowledge of how to warm \texttt{DataPut} (\eg the call to \texttt{DataPut.warm()} in line ~\ref{freshen:datawarm} of Algorithm~\ref{freshen}). If \freshen has started freshening the resource (indicated by the state {\tt running}), both wrapper functions wait for the \freshen thread to finish before returning (line \ref{freshfetch:wait} in Algorithm \ref{freshfetch} and line \ref{freshwarm:wait} in Algorithm \ref{freshwarm}). Finally, if \freshen either did not run or is executing slower than $\lambda$, the wrapper can perform the freshen action itself (line \ref{freshfetch:fetch} Algorithm \ref{freshfetch} and line \ref{freshwarm:warm} in Algorithm \ref{freshwarm}). Not included for brevity in Algorithm \ref{freshen} are the checks to see if the resources have already been freshened by wrapper functions invoked by $\lambda$.

\begin{algorithm}
\caption{Freshen Fetch Function}\label{freshfetch}
\begin{algorithmic}[1]
\State \textbf{Runtime List: } $fr\_state$
\Procedure{$FrFetch$}{$id, \, code$}
\If{fr\_state{[$id$]} $== finished$}
\State \textbf{return} $fr\_state{[id]}.result$\label{freshfetch:return1}
\ElsIf{fr\_state{[$id$]} $== running$}
\State \textbf{FrWait(} $id$ \textbf{)}\label{freshfetch:wait}
\State \textbf{return} $fr\_state{[id]}.result$
\Else
\State fr\_state{[$id$]} $= running$
\State fr\_state{[$id$]}.result = \textbf{Execute(} $code$ \textbf{)}\label{freshfetch:fetch}
\State fr\_state{[$id$]} $= finished$
\State \textbf{return} $fr\_state{[id]}.result$
\EndIf
\EndProcedure
\end{algorithmic}
\end{algorithm}

\begin{algorithm}
\caption{Freshen Warm Function}\label{freshwarm}
\begin{algorithmic}[1]
\State \textbf{Runtime List: } $fr\_state$
\Procedure{$FrWarm$}{$id, \, resource$}
\If{fr\_state{[$id$]} $== finished$}
\State \textbf{return}\label{freshwarm:return1}
\ElsIf{fr\_state{[$id$]} $== running$}
\State \textbf{FrWait(} $id$ \textbf{)}\label{freshwarm:wait}
\State \textbf{return}\label{freshwarm:return2}
\Else
\State fr\_state{[$id$]} $= running$
\State $resource.warm()$ \label{freshwarm:warm}
\State fr\_state{[$id$]} $= finished$
\State \textbf{return}
\EndIf
\EndProcedure
\end{algorithmic}
\end{algorithm}

\tightparagraph{Billing and accounting}
Since \freshen runs in order to benefit the serverless application, the serverless application owner should pay for it. However, as outlined above, \freshen would ideally be triggered based on predictions by the serverless framework. What happens if the platform mispredicts a function call? Confidence in prediction could be used to dictate if \freshen is called or not. Metrics kept inside a container, or communicated to the serverless global scheduling entity, could be used to stop \freshen from running if predictions have been too inaccurate. 
Service categories chosen by the application developer could also control \freshen behavior: aggressive \freshen invocation would be appropriate for latency-sensitive applications. \freshen could be disabled for latency-insensitive functions.
Last, we note providers may be incentivized to offer \freshen because it allows them a way to
monetize warmed containers that are otherwise sitting idle. 

\tightparagraph{Preventing abuse and misconfiguration}
A danger if the application developer were allowed to implement their own \freshen is that the application developer would try to implement their entire function in the \freshen function. This is undesirable and unprofitable for the developer for several reasons: \freshen has no access to function arguments, the application developer is paying for the compute and network resources regardless, and the application would have to handle spurious invocations (mispredictions) gracefully. 

\section{Evaluation}
\label{sec:eval}

\begin{wrapfigure}{R}{0.5\columnwidth}
\begin{center}
\includegraphics[width=.5\columnwidth]{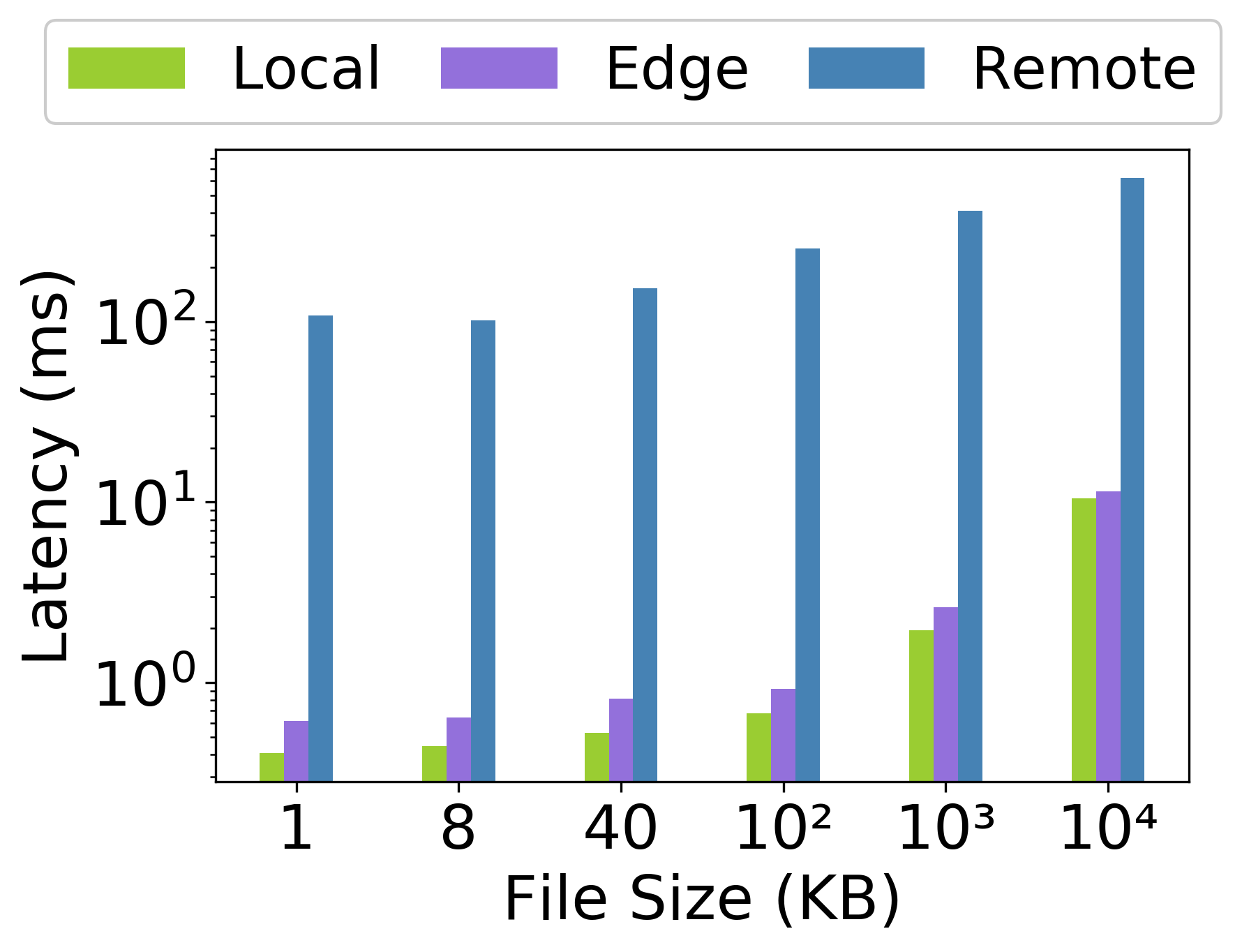}
\end{center}
\caption{\label{fig:filecache} File retrieval overheads to save with \freshen}
\end{wrapfigure}
This section explores the advantages a {\tt freshen} framework could provide. First, the benefits of file caching are evaluated and then improvements from connection warming are illustrated.

\tightparagraph{File caching evaluation} Figure \ref{fig:filecache} demonstrates the potential benefits of proactive file retrieval (file caching). In this benchmark, an OpenWhisk serverless function queries a server for a file of one of six different sizes (x-axis) over a TCP connection. The time measured (y-axis, log scale) is the duration from connection to when the file has been completely received. The file server is located in one of three locations: local on-host (green), edge on-site (purple), and remote off-site (blue). On-site resides on the same 10 Gbps LAN and off-site averages 50ms away. The experiment was conducted using CloudLab \cite{cloudlab} with 20 iterations. The results show how much execution time {\tt freshen} could save a serverless function if \freshen is proactively run. Maximum benefits range from 11-622ms.

\begin{figure}[ht]
\begin{minipage}[b]{0.48\columnwidth}
\centering
\includegraphics[width=1\linewidth, height =0.7\linewidth]{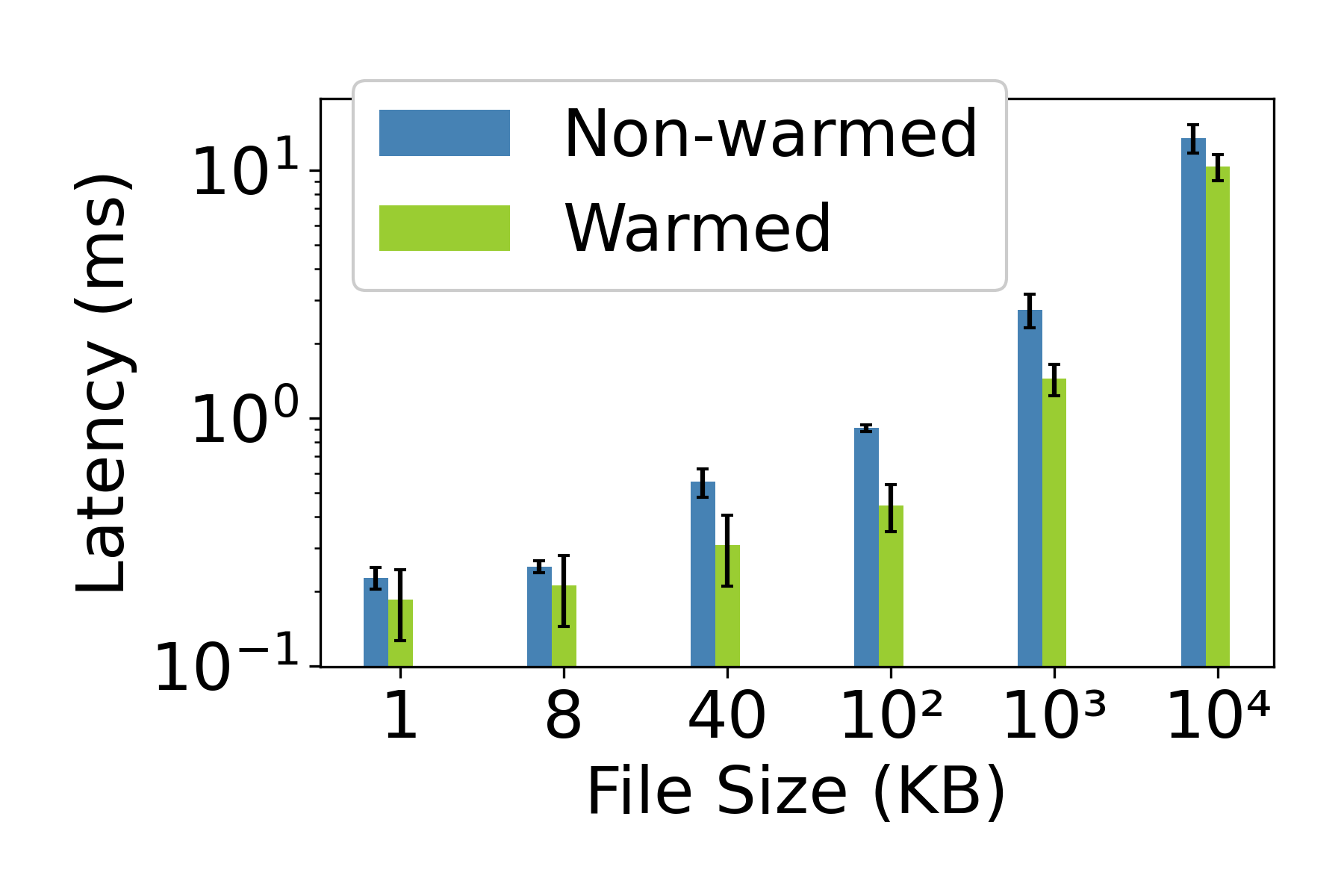}
 
    \captionsetup{font={small}}
   \captionof{figure}{ Warming to cloud}
   
    \label{fig:cwndcloud}
    
\end{minipage}
\begin{minipage}[b]{0.48\columnwidth}
\centering
\includegraphics[width=1\linewidth, height =0.7\linewidth]{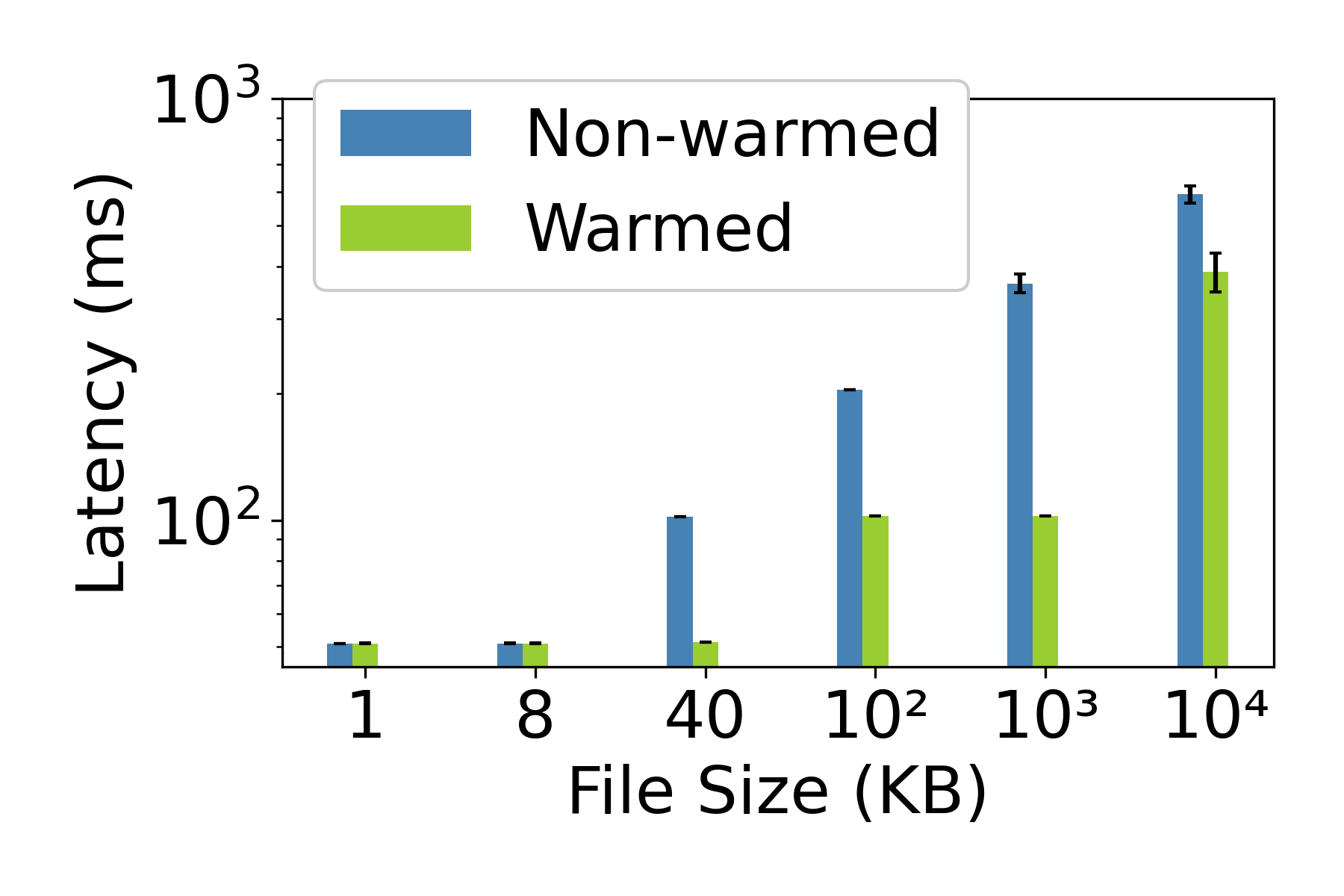}
\captionsetup{font={small}}
\caption{ Warming to edge}
    \label{fig:cwndedge}
\end{minipage}
\end{figure}

\tightparagraph{Warmed connection comparison}
To demonstrate the benefits of {\tt freshen} warming a TCP connection, we run an OpenWhisk serverless function on CloudLab which sends different file sizes to a server. We measure the time of a client initiating a file transfer to the response from the server indicating completion. To understand the potential benefits, we emulate a warmed TCP connection by sending a large file before sending our desired file size. The server is located at two locations, on the same cloud or at the edge (\textasciitilde{}50ms away). The experiment was conducted over 20 iterations. The cloud case is presented in Figure \ref{fig:cwndcloud} and the edge case is presented in Figure \ref{fig:cwndedge}. With smaller file sizes, the performance of warmed and non-warmed is similar.  As file sizes grow, the benefit of warmed connection ranges from 51.22\% to 71.94\%. The edge performance is better because network delay, and not system overheads, dominate totals.

\section{Related Work}
\label{sec:related}
A large body of research focuses on reducing cold start costs. We partition these works into two categories: those that do not propose radical changes to serverless architecture, and those that do. Of those that are compatible with existing serverless infrastructure, techniques include cold start avoidance (runtime reuse), light-weight isolation mechanisms \cite{sock}, caching resources ranging from libraries \cite{sock} to virtual Ethernet infrastructure \cite{agile_cold_starts}, and intelligent host scheduling \cite{fnsched}. Our work has a different focus, optimizing warm starts, but is compatible with these techniques. Works that focus on avoiding cold starts by predicting function execution \cite{serverless_in_the_wild, gunasekaran2020fifer, fnsched} motivate our design because \freshen would be most effective when function invocations are predicted. Of works that propose fundamental changes to serverless architecture such as running more than one function within the same isolation context \cite{sand} or adding distributed application state and/or message passing abilities between serverless functions \cite{sand, cloudburst}, the motivation for \freshen remains but implementation strategies would vary. 

Last, Containerless \cite{containerless} avoids the cost of strong isolation mechanisms
by transforming JavaScript serverless functions into Rust via dynamic tracing. Their 
dynamic tracing design, as well as analysis of the resulting traces, could help inform how \freshen could be inferred by providers.

\section{Discussion and Conclusion}
\label{sec:discuss}
\tightparagraph{Discussion} There exists many opportunities for future work. First, the system should be fully deployed and thoroughly evaluated. Quantifying how \freshen affects variability in application behavior would be an important component of this evaluation. Prediction success must be additionally quantified, especially in the case of non-deterministic function chains.
In addition, the framework must be analyzed for misuse and hardened as necessary. Impact on developer burden, or the extent to which providers can automatically generate \freshen must also be further studied. Finally, integrating \freshen into serverless architectures that provide different isolation scopes is an additional area for future study (\eg Azure offers chain-level isolation).

\tightparagraph{Conclusion} This paper proposes a new primitive to serverless language runtimes called \freshen. With \freshen, developers or service providers specify functionality to complete before a given function executes. This proactive framework allows for overheads associated with serverless functions to be mitigated at execution time, which improves function responsiveness. We argue predictive opportunities exist to enable \freshen to be run with ample time. A high-level design and implementation are presented, along with preliminary results to show potential benefits of the scheme.

\bibliographystyle{plain}
\bibliography{refs,related_work}

\begin{thebibliography}{10}

\bibitem{sand}
Istemi~Ekin Akkus, Ruichuan Chen, Ivica Rimac, Manuel Stein, Klaus Satzke,
  Andre Beck, Paarijaat Aditya, and Volker Hilt.
\newblock {SAND}: Towards high-performance serverless computing.
\newblock In {\em 2018 {USENIX} Annual Technical Conference ({USENIX} {ATC}
  18)}, pages 923--935, Boston, MA, July 2018. {USENIX} Association.

\bibitem{cloudlab}
Dmitry Duplyakin, Robert Ricci, Aleksander Maricq, Gary Wong, Jonathon Duerig,
  Eric Eide, Leigh Stoller, Mike Hibler, David Johnson, Kirk Webb, Aditya
  Akella, Kuangching Wang, Glenn Ricart, Larry Landweber, Chip Elliott, Michael
  Zink, Emmanuel Cecchet, Snigdhaswin Kar, and Prabodh Mishra.
\newblock The design and operation of {CloudLab}.
\newblock In {\em Proceedings of the {USENIX} Annual Technical Conference
  (ATC)}, pages 1--14, July 2019.

\bibitem{gunasekaran2020fifer}
Jashwant~Raj Gunasekaran, Prashanth Thinakaran, Nachiappan Chidambaram,
  Mahmut~T. Kandemir, and Chita~R. Das.
\newblock Fifer: Tackling underutilization in the serverless era.
\newblock arXiv, 2020.

\bibitem{containerless}
Emily Herbert and Arjun Guha.
\newblock A language-based serverless function accelerator.
\newblock arXiv, 2019.

\bibitem{keshav1995packet}
Srinivasan Keshav.
\newblock Packet-pair flow control.
\newblock {\em IEEE/ACM transactions on Networking}, pages 1--45, 1995.

\bibitem{Mace:2018:UCP:3190508.3190526}
Jonathan Mace and Rodrigo Fonseca.
\newblock Universal context propagation for distributed system instrumentation.
\newblock In {\em Proceedings of the Thirteenth EuroSys Conference}, EuroSys
  '18, pages 8:1--8:18, New York, NY, USA, 2018. ACM.

\bibitem{agile_cold_starts}
Anup Mohan, Harshad Sane, Kshitij Doshi, Saikrishna Edupuganti, Naren Nayak,
  and Vadim Sukhomlinov.
\newblock Agile cold starts for scalable serverless.
\newblock In {\em 11th {USENIX} Workshop on Hot Topics in Cloud Computing
  (HotCloud 19)}, Renton, WA, July 2019. {USENIX} Association.

\bibitem{sock}
Edward Oakes, Leon Yang, Dennis Zhou, Kevin Houck, Tyler Harter, Andrea
  Arpaci-Dusseau, and Remzi Arpaci-Dusseau.
\newblock {SOCK}: Rapid task provisioning with serverless-optimized containers.
\newblock In {\em 2018 {USENIX} Annual Technical Conference ({USENIX} {ATC}
  18)}, pages 57--70, Boston, MA, July 2018. {USENIX} Association.

\bibitem{serverless_in_the_wild}
Mohammad Shahrad, Rodrigo Fonseca, Inigo Goiri, Gohar Chaudhry, Paul Batum,
  Jason Cooke, Eduardo Laureano, Colby Tresness, Mark Russinovich, and Ricardo
  Bianchini.
\newblock Serverless in the wild: Characterizing and optimizing the serverless
  workload at a large cloud provider.
\newblock In {\em 2020 {USENIX} Annual Technical Conference ({USENIX} {ATC}
  20)}, pages 205--218. {USENIX} Association, July 2020.

\bibitem{cloudburst}
Vikram Sreekanti, Chenggang Wu, Xiayue~Charles Lin, Johann Schleier-Smith,
  Joseph~E. Gonzalez, Joseph~M. Hellerstein, and Alexey Tumanov.
\newblock Cloudburst.
\newblock {\em Proceedings of the VLDB Endowment}, 13(12):2438–2452, Aug
  2020.

\bibitem{fnsched}
Amoghvarsha Suresh and Anshul Gandhi.
\newblock Fnsched: An efficient scheduler for serverless functions.
\newblock In {\em Proceedings of the 5th International Workshop on Serverless
  Computing}, WOSC '19, page 19–24, New York, NY, USA, 2019. Association for
  Computing Machinery.

\bibitem{sequoia}
Ali Tariq, Austin Pahl, Sharat Nimmagadda, Eric Rozner, and Siddharth Lanka.
\newblock Sequoia: Enabling quality-of-service in serverless computing.
\newblock In {\em Proceedings of the Annual Symposium on Cloud Computing
  (SoCC)}, 2020.

\bibitem{curtains}
Liang Wang, Mengyuan Li, Yinqian Zhang, Thomas Ristenpart, and Michael Swift.
\newblock Peeking behind the curtains of serverless platforms.
\newblock In {\em 2018 {USENIX} Annual Technical Conference ({USENIX} {ATC}
  18)}, pages 133--146, Boston, MA, 2018. {USENIX} Association.

\end{thebibliography}

\end{document}